\documentclass[preprint, 3p, twocolumn]{elsarticle}

\usepackage{hyperref}
\usepackage{array, multirow} 
\usepackage{amssymb} 
\usepackage{amsthm} 
\usepackage{amsmath}
\usepackage{textcomp} 
\usepackage[table]{xcolor}

\journal{Earth and Planetary Science Letters}

\biboptions{authoryear}

\begin{document}


\begin{frontmatter}

\title{Importance of the advection scheme for the simulation of water isotopes over Antarctica by atmospheric general circulation models: a case study for present-day and Last Glacial Maximum with LMDZ-iso}


\author[AWI]{A. Cauquoin\corref{correspondingauthor}}
\cortext[correspondingauthor]{Corresponding author}
\ead{alexandre.cauquoin@awi.de}

\author[LMD]{C. Risi}

\author[LTE]{{\'E}. Vignon}

\address[AWI]{Alfred Wegener Institute, Helmholtz Centre for Polar and Marine Sciences, Bremerhaven, Germany}
\address[LMD]{Laboratoire de M{\'e}t{\'e}orologie Dynamique/Institut Pierre-Simon Laplace (LMD/IPSL), CNRS, Sorbonne Universit{\'e}s, UPMC Univ Paris 06, Paris, France}
\address[LTE]{Environmental Remote Sensing Laboratory (LTE), {\'E}cole Polytechnique F{\'e}d{\'e}rale de Lausanne (EPFL), Lausanne, Switzerland}

\begin{abstract}
Atmospheric general circulation models (AGCMs) are known to have a warm and isotopically enriched bias over Antarctica. We test here the hypothesis that these biases are partly consequences of a too diffusive advection. Exploiting the LMDZ-iso model, we show that a less diffusive representation of the advection, especially on the horizontal, is very important to reduce the bias in the isotopic contents of precipitation above this area. The choice of an appropriate representation of the advection is thus essential when using GCMs for paleoclimate applications based on polar water isotopes. Too much diffusive mixing along the poleward transport leads to overestimated isotopic contents in water vapor because dehydration by mixing follows a more enriched path than dehydration by Rayleigh distillation. The near-air surface temperature is also influenced, to a lesser extent, by the diffusive properties of the advection scheme directly via the advection of the air and indirectly via the radiative effects of changes in high cloud fraction and water vapor. A too diffusive horizontal advection increases the temperature and so also contributes to enrich the isotopic contents of water vapor over Antarctica through a reduction of the distillation. The temporal relationship, from Last Glacial Maximum (LGM) to present-day conditions, between the mean annual near-air surface temperature and the water isotopic contents of precipitation for a specific location can also be impacted, with significant consequences on the paleo-temperature reconstruction from observed changes in water isotopes.
\end{abstract}


\end{frontmatter}


\section{Introduction}

Water stable isotopologues (hereafter designated by the term ``water isotopes''), are integrated tracers of the water cycle. Especially, the isotopic composition recorded in polar ice cores enabled the reconstruction of past temperature variations \citep[and references therein]{jouzel2013}. For example, low accumulation sites that are typical on the East Antarctic Plateau ($<$ 10~cm water-equivalent yr$^{-1}$) provided the longest ice core records, making it possible to reconstruct past climate over several glacial-interglacial cycles  \citep{jouzel2007}. However, the interpretation of isotope signals remains challenging because of the numerous and complex processes involved (water vapor transport, fractionation during the phase changes in the water cycle, distillation effect\ldots). This is particularly the case for Antarctica because this part of the world is subject to extreme weather conditions.

To improve our knowledge on the mechanisms controlling the water isotopes distribution, atmospheric general circulation models (AGCMs) enhanced by the capability to explicitly simulate the hydrological cycle of the water isotopes (H$_2$$^{16}$O, HDO, H$_2$$^{17}$O, H$_2$$^{18}$O) are now frequently used \citep{joussaume1984, risi2010_lmdz, werner2011}. Water isotopes in climate models have been used, for example, to better understand how the climatic signal is recorded by isotopes in polar ice cores at paleoclimatic time scales \citep{werner2001}. 

However, some issues remain concerning the simulation of the climate over the Antarctic continent by AGCMs. For example, they frequently present a near-surface warm bias over this area \citep{masson-delmotte2006} and isotopic values in precipitation that are not depleted enough compared to observations \citep{lee2007, risi2010_lmdz, werner2011}. This raises the question why many of the AGCMs have these warm and enriched in heavy water isotopes biases over Antarctica. 

In this paper, we hypothesize that one part of these biases is associated with an excessively diffusive water vapor transport, i.e.\ transport that is associated with too much mixing. According to previous studies, the diffusive properties of the advection scheme in the AGCMs, on the horizontal as well on the vertical, can have an impact on the simulation of humidity and of its water isotope contents. On the horizontal, dehydration of air masses by mixing with a drier air mass leads to more enriched water vapor than dehydration by condensation and associated Rayleigh distillation \citep{galewsky2010}. For the same reason, poleward water vapor transport by eddies (which act as mixing) leads to more enriched water vapor in Antarctica than transport by steady advection \citep{hendricks2000}. On the vertical, the excessive diffusion during water vapor transport seems to be the cause of the moist bias found in most AGCMs in the tropical and subtropical mid and upper troposphere, and of the poor simulation of isotopic seasonality in the subtropics \citep{risi2012}. The diffusivity of the advection scheme in the vertical has also important consequences on modeling of tracers like tritium by affecting greatly its residence time in the stratosphere, and so its downward transport from the stratosphere to the troposphere \citep{cauquoin2016_jgr}.

The goal of this paper is to test whether the warm and enriched biases in Antarctica are associated with an excessively diffusive water vapor transport, both on the horizontal and on the vertical. The diffusive character of the advection can be varied by modifying either the advection scheme or the resolution of the simulation, and we test both possibilities. Finally, we explore if a too diffusive water vapor transport can affect the temporal water isotopes -- temperature slope between the Last Glacial Maximum (LGM, 21~ka) and present-day periods.

\section{Model, simulations and data}
\subsection{Model and simulations}

We use here the isotopic AGCM LMDZ-iso \citep{risi2010_lmdz} at a standard latitude-longitude R96 grid resolution (2.5$^{\circ}$ $\times$ 3.75$^{\circ}$), and with 39 layers in the vertical spread in a way to ensure a realistic description of the stratosphere and of the Brewer-Dobson circulation \citep{lott2005}. Water isotopes are implemented in a way similar to other state-of-the-art isotope-enabled AGCM \citep{risi2010_lmdz}. The isotopic composition of glaciers $R_{\textrm{glacier}}$ is calculated prognostically in the model. It is a precipitation-weighted average of the previous snow fall. At each time step and in each grid box, it is updated as: 
\begin{equation}
\begin{split}
& R_{\textrm{glacier}}(t+dt)  \\
& =	\frac{ h_{\textrm{glacier}} \times R_{\textrm{glacier}}(t) + iso_{\textrm{snowfall}} \times dt }{ h_{\textrm{glacier}} + H_2O_{\textrm{snowfall}} \times dt }
\label{eq1}
\end{split}
\end{equation}
with $h_{\textrm{glacier}}=$ 20 kg.m$^{-2}$ the height scale for the glacier, $iso_{\textrm{snowfall}}$ and $H_2O_{\textrm{snowfall}}$ the snowfall of isotopes and standard water in kg.m$^{-2}$.s$^{-1}$ and $dt = 30 \times 60$ s. No fractionation is assumed during the runoff and sublimation of glaciers ice. The model has been validated at global scale for the simulation of both atmospheric \citep{hourdin2006} and isotopic \citep{risi2010_lmdz} variables, and has been extensively compared to various isotopic measurements in polar regions \citep{casado2013, steen-larsen2013, steen-larsen2014_cp, steen-larsen2017, bonne2014, bonne2015, touzeau2016, ritter2016, stenni2016}. LMDZ-iso is also able to simulate the H$_2$$^{17}$O distribution \citep{risi2013} but we do not consider it here because the limitations inherent to the AGCMs lead to strong uncertainties and numerical errors on the spatio-temporal distribution of this isotope. 2 years of spin-up have been performed for all the simulations presented hereafter. 

To quantify the effects of the prescribed advection scheme on water stable isotope values over Antarctica, we first performed three sensitivity simulations with LMDZ-iso under present-day conditions for the post spin-up period 1990-2008 (i.e.\ 19 model years), following the model setup from \citep{cauquoin2016_jgr} i.e.\ simulations follow the AMIP protocol \citep{gates1992}, forced by monthly observed sea-surface temperatures and nudged by the horizontal winds from 20CR reanalyses \citep{compo2011}: (1) one control simulation with the \citet{vanleer1977} advection scheme (called VL), which is a second order monotonic finite volume scheme prescribed by default in the standard version of the model \citep{risi2010_lmdz}; and two other simulations whose the van Leer advection scheme has been replaced by a single upstream scheme \citep{godunov1959} on (2) the horizontal plane (UP\_xy) and on (3) the vertical direction (UP\_z). Depending on one tunable parameter, the LMDZ model can be used with these 2 versions of the advection scheme according to the object of study \citep{risi2012}. The advection scheme in the simulations presented in the LMDZ-iso reference paper from \citet{risi2010_lmdz} was set erroneously to the simple upstream scheme rather than to the van Leer's scheme \citep{risi2010_corr}, and has little influence on their simulated spatial and temporal distributions of water isotopes at a global scale. However, as we will show here, this has considerable effect on the spatial distribution of these proxies over region with extreme weather conditions such as Antarctica. The 2-year spin-up time is enough to reach equilibrium. In the VL simulation, the globally and annually average values of temperature and $\delta^{18}$O in precipitation for 1990 are $13.06$$^{\circ}$C and $-7.53$ \textperthousand\ respectively, very close to the average values over the whole period 1990--2008 ($13.14$$^{\circ}$C and $-7.57$ \textperthousand) within the average interannual variability of $0.15$$^{\circ}$C and $0.06$ \textperthousand. The conclusion is the same if we focus on the 60$^{\circ}$S--90$^{\circ}$S area instead: the average values of temperature and $\delta^{18}$O in precipitation for the year 1990 are $-17.89$$^{\circ}$C and $-25.45$ \textperthousand\ respectively, very close to the average values over the whole period 1990--2008 ($-17.71$$^{\circ}$C and $-25.34$ \textperthousand) within the average interannual variability of $0.43$$^{\circ}$C and $0.29$ \textperthousand.

The detailed description of the mixing ratio by the van Leer's (\citeyear{vanleer1977}) scheme and its comparison with the upstream scheme  \citep{godunov1959} can be found in the Appendix A of \citet{cauquoin2016_jgr}. To resume, the mixing ratio at the left boundary of box $i$, $q_{i-1/2}$, is calculated as a linear combination of the mixing ratio in the boxes $i-1$ and $i$ in the van Leer's scheme whereas in the upstream scheme $q_{i-1/2} = q_{i-1}$. This means that in the upstream scheme, even if the air mass flux from grid box $i-1$ to grid box $i$ is very small, the air that is advected into box $i$ has the same water vapor mixing ratio as grid box $i-1$ as a whole. This makes the upstream scheme much more diffusive. An example of the effect of these 2 different advection schemes in a very idealized case is given in the figure 3 (top panel) of \citet{hourdin1999}. In a single dimension case, a tracer distribution is initially rectangle and is advected with a constant velocity such that Courant number is equal to 0.2. After 70 iterations, the initial rectangle shape is almost unchanged with the van Leer scheme, whereas it spreads into a flat Gaussian shape with the upstream scheme. Quantitatively, almost half of the tracer mass becomes outside the initial rectangle shape with the upstream scheme.

Increasing the grid resolution is equivalent to using an advection scheme that is less diffusive. Indeed, these finite-difference schemes are discretization methods and so depend on the chosen spatial resolution. To check that our findings and conclusions are consistent, we performed two more UP\_xy and VL simulations under present-day conditions with the same configuration as presented above but at the R144 resolution (latitude-longitude grid resolution of 1.27$^{\circ}$ $\times$ 2.5$^{\circ}$).

Finally, we evaluate the impacts of applying advection schemes with different diffusive properties on the temporal relationship between water isotopic contents of precipitation and mean air surface temperature, essential for paleo-temperature reconstructions. For this, we effected 12 years long (the two first years being used for the spin-up) UP\_xy and VL simulations for present-day ``free'' (i.e.\ not nudged by the 20CR reanalyses, still following the AMIP protocol) and LGM conditions at the R96 resolution. For the LGM simulations, the PMIP3 protocol is applied \citep{braconnot2012}. Orbital parameters and greenhouse gas concentrations are set to their LGM values. ICE-5G ice sheet conditions are applied \citep{peltier1994}. LMDZ-iso is forced by the climatological sea-surface temperatures (SSTs) and sea ice from the IPSL-CM4 model (Marti et al., 2005). The SST simulated by IPSL-CM4 for pre-industrial conditions (PI) has a global bias of -0.95 Kelvin with a cold bias in the mid-latitudes, a warm bias on the eastern side of the tropical oceans and on the Southern Ocean, and a particularly strong cold bias in the North Atlantic \citep{hourdin2013}. To avoid confusing these biases with LGM -- present-day signals, we use the SSTs from an IPSL PI simulation in the following way to cancel out the biases in the IPSL model common to both the LGM and PI simulations (see details in \citet{risi2010_lmdz}): SST forcing $=$ SST$_{\textrm{LGM}}$ $-$ SST$_{\textrm{PI}}$ $+$ SST$_{\textrm{AMIP}}$. We set the sea surface $\delta^{18}$O to $+1.2$ \textperthousand\ and assume no glacial change in the mean deuterium excess in the ocean, as in \citet{risi2010_lmdz}. Note once again that the 2-year spin-up time is enough to reach equilibrium. In the VL present-day ``free'' simulation, the globally and annually averaged values of temperature and $\delta^{18}$O in precipitation for the first post spin-up model year are $12.54$$^{\circ}$C and $-8.27$ \textperthousand\ respectively, very close to the average values over the 10 years of simulation ($12.54$$^{\circ}$C and $-8.34$ \textperthousand) in comparison to the interannual variabilities of $0.06$$^{\circ}$C and $0.06$ \textperthousand. The conclusion still holds if we focus on the 60$^{\circ}$S--90$^{\circ}$S area instead: the average values of temperature and $\delta^{18}$O in precipitation for the first post spin-up year are $-22.03$$^{\circ}$C and $-26.08$ \textperthousand\ respectively, very close to the average values over the 10 years of simulation ($-22.04$$^{\circ}$C and $-26.04$ \textperthousand) within the average interannual variability of $0.10$$^{\circ}$C and $0.21$ \textperthousand.

We express the isotopic composition of difference water bodies in the usual $\delta$-notation as the deviation from the Vienna Standard Mean Ocean Water (V-SMOW). So for H$_2$$^{18}$O, the $\delta^{18}$O value is calculated as $\delta^{18}$O $=$ (([H$_2$$^{18}$O]$/$[H$_2$$^{16}$O])$_{\textrm{sample}}$ $/$ ([H$_2$$^{18}$O]$/$[H$_2$$^{16}$O])$_{\textrm{V-SMOW}}$ $-$ 1) $\times$ 1000. Long-time mean $\delta$ values are then calculated as precipitation-weighted mean. For the quantitative model-data comparisons, we retrieve the model values at data geographical coordinates by bilinear interpolation. Without such an interpolation, i.e.\ considering the nearest grid point instead, the root-mean-squared errors (RMSE) of mean $\delta^{18}$O and temperature from the VL simulation differ only by 0.5 \textperthousand\ and 0.14$^{\circ}$C compared to the results presented below, so the uncertainty associated with the model-data co-location is small.

\subsection{Observations} \label{data}
To analyze the model performance over Antarctica under present-day conditions, we make use of the observational isotope database compiled by \citet{masson-delmotte2008}. We also focus especially on the East Antarctic plateau (defined by the black bold contour of 2500~m above sea level elevation in Figure \ref{fig1}) because this area provides the main reconstructions of past climate based on the interpretation of water stable isotope records. To compare the model-data agreement of our simulations with van Leer and upstream advection schemes for atmospheric boundary layer, inversion temperatures, we use additional datasets for the EPICA Dome C station (EDC: [75.10$^{\circ}$ S; 123.35$^{\circ}$ E]): the surface temperature, the 10m-temperature and the downward longwave radiative flux at the surface (LW$_{\textrm{dn}}$) over the period 2011--2018 thanks to the CALVA program and the Baseline Surface Radiation Network (BRSN) \citep[and references therein]{vignon2018}; and the observed precipitable water from radiosoundings data for the period 2010--2017. We are aware that the observations periods are not the same than our model period (1990--2008), giving a possible bias in the model-data comparison. For the cloud cover, we use the CALIPSO--GOCCP observations \citep{chepfer2010} over the period 2007--2008. We compare these data with the high-level cloud fraction over EDC simulated by LMDZ-iso through the COSP package \citep{bodas-salcedo2011}, which allows to simulate what the CALIOP instrument on-board CALIPSO would measure if it was into orbit above the simulated atmosphere.

\section{Results and discussion}
\subsection{Model-data comparison for present-day conditions}
\subsubsection{Water stable isotopes}

\begin{figure*}[htbp]
\center
\includegraphics[width=0.95\textwidth]{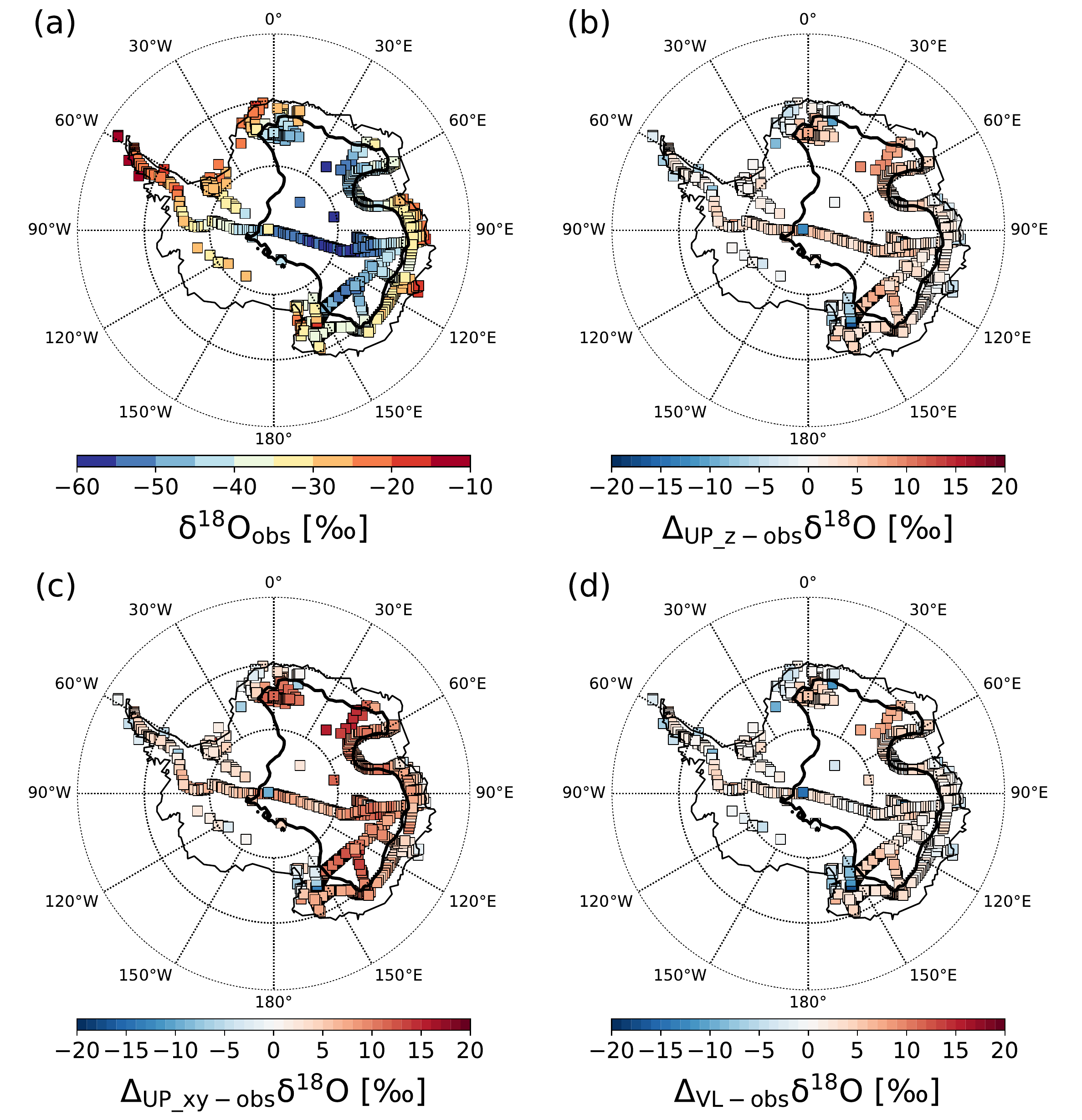} 
\caption{Map of Antarctica showing (a) the observed $\delta^{18}$O values from the compilation by \citet{masson-delmotte2008}, (b) the difference between the simulated $\delta^{18}$O in precipitation and the $\delta^{18}$O observations for the UP\_z, (c) UP\_xy and (d) VL simulations. The bold black line shows the contour of 2500~m above sea level elevation.}
\label{fig1}
\end{figure*}

Figure 1 shows the observed annual mean $\delta^{18}$O values in the snow surface in Antarctica compiled by \citet{masson-delmotte2008} (Figure \ref{fig1}a) and the difference with the modeled annual $\delta^{18}$O in precipitation from the UP\_z (Figure \ref{fig1}b), UP\_xy (Figure \ref{fig1}c) and VL (Figure \ref{fig1}d) simulations. The spatial average over the 60$^{\circ}$S--90$^{\circ}$S area of standard deviations of long-time $\delta^{18}$O and $\delta$D model values is of 0.97 and 7.48 \textperthousand, respectively. The results from the VL simulation are in better agreement with the $\delta^{18}$O observations over Antarctica (Figure \ref{fig1}d). This is confirmed by the smaller root-mean-squared error of modeled $\delta^{18}$O in precipitation from the VL simulation, calculated as the difference between the observed annual mean values and the LMDZ-iso results (RMSE = 4.47 \textperthousand, i.e.\ 12.2 \% of the observed mean Antarctic $\delta^{18}$O value). The results from the VL simulation for the other isotopic variable $\delta$D is also the closest of the observations with a RMSE of 40.93 \textperthousand\ (Table \ref{table1}, red background). Our simulated $\delta^{18}$O in precipitation is very sensitive to the choice of the advection scheme on the horizontal plane, with more enriched values when a more diffusive advection scheme is applied (Figure \ref{fig1}c). This is reflected by the mean $\delta^{18}$O model value and RMSE, higher by 3.16 and 4.31 \textperthousand\ than the VL simulation values. On the contrary, the sensitivity of Antarctic $\delta^{18}$O values to the diffusive properties of vertical advection is weak (Figure \ref{fig1}b) with an RMSE very close of the VL one (4.84 \textperthousand). The results from the VL simulation for the other isotopic variable $\delta$D is also the closest of the observations with a RMSE of 40.93 \textperthousand\ (Table \ref{table1}, red background). According to the observations, the East Antarctic plateau is where the water isotope values are the lowest (mean $\delta^{18}$O below $-40$ \textperthousand, Figure \ref{fig1}a) due to the very low temperatures. Because of the extreme cold and dry conditions at this area, one can see that the main disagreements between model outputs and observations are located at this place (Figure \ref{fig1} and blue background of Table \ref{table1}). Again, the isotopic outputs from the VL simulation are in better agreement with the observations (Table \ref{table1}, blue background). These first results confirm that an excessively diffusive water vapor transport influences significantly the simulated isotopic and temperature values over Antarctica.

\begin{table*}[htbp]
\caption{Observed and simulated annual mean values of temperature ($T$), $\delta^{18}$O and $\delta$D for the full Antarctic dataset (red background) and restricted to the East-Antarctic plateau (blue background), and the corresponding RMSE.}
\centering
\begin{tabular}{l | c | cc | cc | cc}
\hline
							&	Mean	&	Mean	& RMSE		&	Mean	& RMSE		&	Mean	&	RMSE	\\
							&	data		&	UP\_z	& UP\_z		&	UP\_xy	& UP\_xy		&	VL		&	VL		\\
\hline
\rowcolor{red!25}
$T$ ($^{\circ}$C)				&	$-36.93$	&	$-30.51$	&	$7.50$	&	$-30.69$	&	$7.31$	&	$-31.54$	&	$6.60$	\\
\rowcolor{red!25}
$\delta^{18}$O (\textperthousand)	&	$-36.76$	&	$-34.85$	&	$4.84$	&	$-31.43$	&	$7.63$	&	$-35.74$	&	$4.47$	\\
\rowcolor{red!25}
$\delta$D (\textperthousand)		&	$-289.62$	&	$-272.28$	&	$43.76$	&	$-251.34$	&	$62.00$	&	$-279.49$	&	$40.93$	\\
\hline
\rowcolor{blue!25}
$T$ ($^{\circ}$C)				&	$-47.46$	&	$-39.49$	&	$8.38$	&	$-39.88$	&	$7.99$	&	$-40.71$	&	$7.23$	\\
\rowcolor{blue!25}
$\delta^{18}$O (\textperthousand)	&	$-46.77$	&	$-42.27$	&	$5.03$	&	$-37.37$	&	$9.69$	&	$-43.76$	&	$3.80$	\\
\rowcolor{blue!25}
$\delta$D (\textperthousand)		&	$-366.98$	&	$-325.37$	&	$43.79$	&	$-291.99$	&	$76.44$	&	$-336.25$	&	$33.58$	\\
\hline
\end{tabular}
\label{table1}
\end{table*}

\subsubsection{Temperature} \label{section_temp}
The bias in temperature is deteriorated about in the same way when applying a more diffusive advection on the vertical direction or on the horizontal plane, as shown with the RMSE of annual mean temperature of 7.50, 7.31 and 6.60$^{\circ}$C for the UP\_z, UP\_xy and VL simulations respectively (Table \ref{table1}, red background). This tendency is the same when focusing on the East Antarctic plateau. However, in average over the East Antarctic plateau, the temperatures of $-30.51$, $-30.69$ and $-31.54$$^{\circ}$C from the UP\_z, UP\_xy and VL simulations are all within the spatial average of standard deviations of long-time temperature values (0.87$^{\circ}$C). These values are all much warmer than the average observed temperature (-36.93$^{\circ}$C). This shows that other factors than advection are responsible for the warm biais.

\begin{figure*}[htbp]
\center
\includegraphics[width=0.95\textwidth]{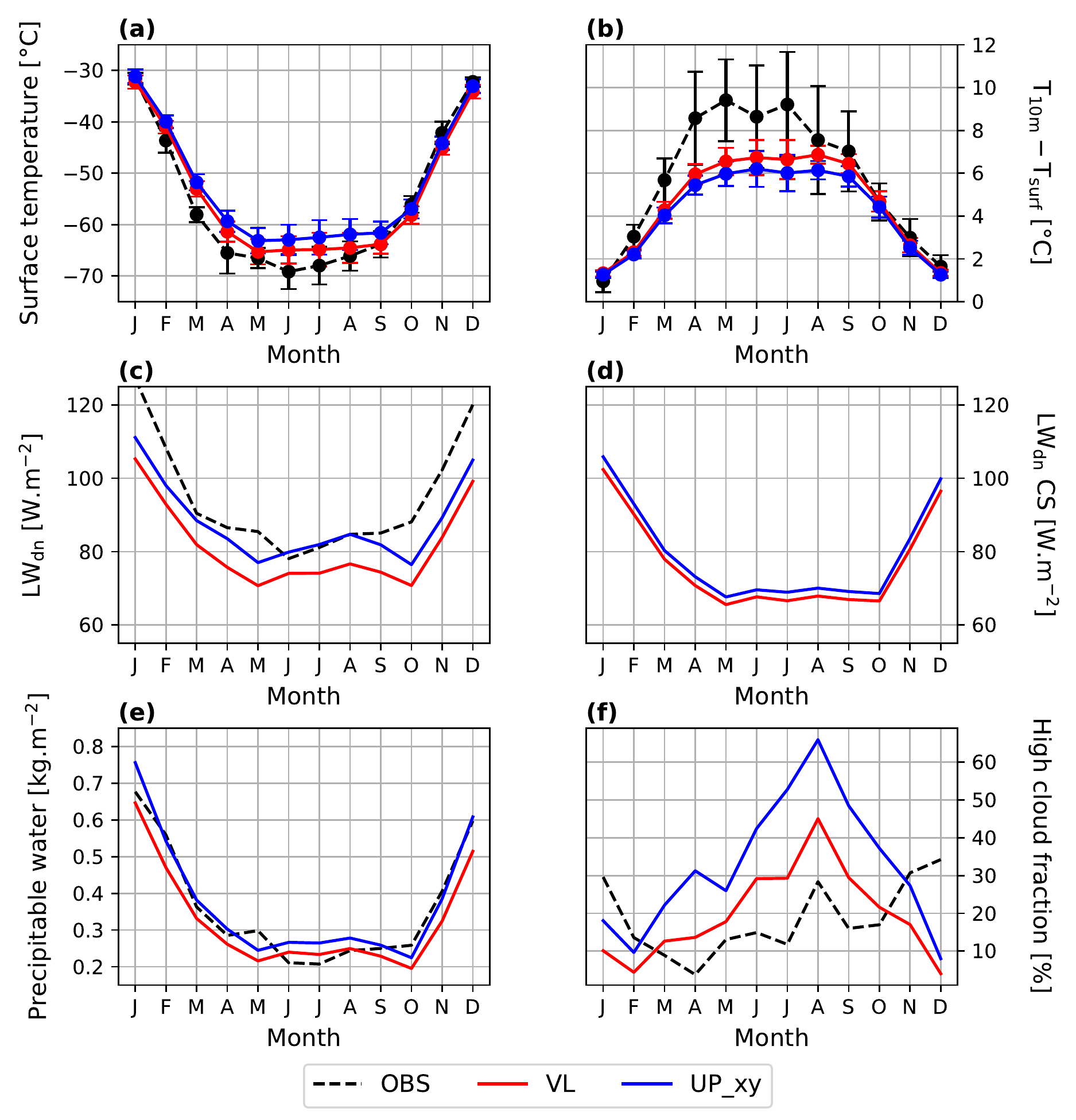} 
\caption{Multi-year monthly mean variations of (a) surface temperature, (b) near-surface thermal inversion (defined as the difference between the 10m-temperature and the surface temperature), (c and d) total and clear-sky component of downward longwave radiative flux at the surface, (e) precipitable water and (f) high cloud fraction above EPICA Dome C. The red and blue curves correspond to the VL and UP\_xy simulation results, respectively. A comparison with observations (see section  \ref{data}) is made when possible (dashed black lines).}
\label{fig2}
\end{figure*}

It has been suggested that the Antarctic warm bias in AGCMs could be linked to the general poor representation of the polar atmospheric boundary layer and related atmospheric inversion temperatures in these models \citep{krinner1997}. \citet{cesana2012} have also shown that CMIP5 models generate too many high-level clouds (i.e.\ above an altitude of 6.72~km), that can partly explain the overestimation of temperatures in Antarctica, due to their effect on downwelling longwave radiation. To go further, we compare the seasonal signals of surface temperature, near-surface thermal inversion (defined here as the difference between the 10m-temperature and the surface temperature), downward longwave radiative flux at the surface (LW$_{\textrm{dn}}$), integrated water vapor column and high cloud fraction at EDC from VL and UP\_xy simulations with meteorological observations (see Figure \ref{fig2} and the description of the data in section \ref{data}). The difference in LW$_{\textrm{dn}}$ between the simulations VL and UP\_xy at EDC is also present all over the East Antarctic plateau, confirming that EDC is representative of this area.

Both VL and UP\_xy simulations have a warm bias at the surface (Figure 2a) and underestimate the near-surface inversion (Figure \ref{fig2}b), especially during the austral winter. This is mostly explained by an overly active turbulent mixing within stable boundary layers in LMDZ5 \citep{vignon2017}. The disagreement with the observed near-surface inversion at EDC is exacerbated in the UP\_xy simulation because the modeled surface temperature is more affected by the change of advection scheme than the modeled 10m-temperature. This is consistent with a higher total LW$_{\textrm{dn}}$ flux over EDC in the UP\_xy simulation than in the VL one (Figure \ref{fig2}c), by around 5 W.m$^{-2}$ in summer and 8 W.m$^{-2}$ in austral winter. The difference of LW$_{\textrm{dn}}$ between our two simulations can be due to two aspects: the fraction of high cloud and the amount of water vapor (e.g., \citet{vignon2018}). The modeled high-level cloud fraction over EDC for the period 2007--2008 is overestimated compared to the CALIPSO--GOCCP observations (Figure \ref{fig2}f). This finding is consistent with the results from \citet{cesana2012} for CMIP5 models and with \citet{lacour2018} for dry areas over Greenland. This result also concurs with an overestimated relative humidity with respect to ice in the troposphere compared to radiosoundings (not shown) during winter seasons, especially in the UP\_xy simulation. The disagreement of high cloud cover is clearly enhanced in the UP\_xy case, which is consistent with stronger LW$_{\textrm{dn}}$ in the UP\_xy simulation. The comparison of LW$_{\textrm{dn}}$ under clear sky in our two simulations (Figure \ref{fig2}d) can give us information about the contribution of cloud and water vapor on the variations of total LW$_{\textrm{dn}}$ (Figure \ref{fig2}c). The stronger simulated LW$_{\textrm{dn}}$ in the UP\_xy simulation is due mainly to the high cloud in winter (70 \%) and to both high cloud and water vapor during summer (42.5 and 57.5 \%, respectively). The effect of water vapor on downward longwave radiative flux is also confirmed by the higher amount of precipitable water over EDC in the UP\_xy simulation (Figure \ref{fig2}e). From these results, we deduce that the warmer surface temperature in the UP\_xy simulation is due to the higher air temperature (direct effect of the horizontal upstream advection) and to the radiative amplification from high clouds and in a lesser extent from water vapor (indirect effect of the horizontal upstream advection).

The near-surface warm bias in LMDZ-iso, which is most pronounced for the coldest temperatures (see Figure \ref{fig3}), has the consequence that the distillation is not strong enough. Some microphysical processes and kinetic fractionation at very low temperature can be missed too. These different aspects could contribute to an overestimation of the $\delta^{18}$O and $\delta$D in precipitation over Antarctica. Finally, the 20CR reanalysis assimilate only surface observations of air pressure and use the observed monthly sea surface temperature and sea ice concentration as lower boundary conditions. These less strong constraints, compared to other reanalyses, may cause biases on the surface temperature over the poles (A. Orsi, personal communication), that can impact our isotopic delta values.

\subsubsection{Spatial $\delta^{18}$O--temperature relationship} \label{section_spatial_rel}
\begin{figure}[htb]
\center
\includegraphics[width=0.95\columnwidth]{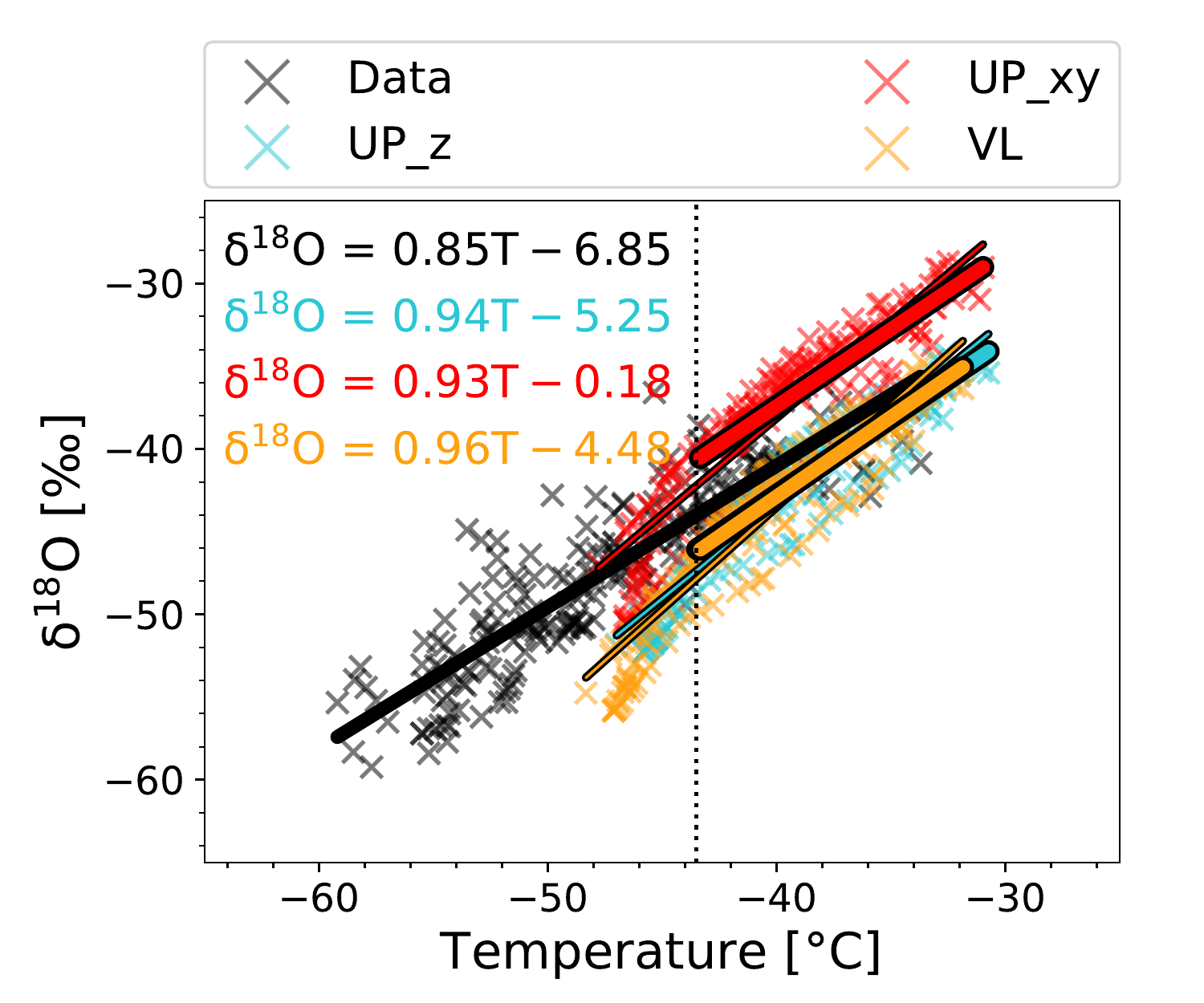} 
\caption{Relationship between $\delta^{18}$O and temperature on the East-Antarctic plateau according to the observations (black) and the UP\_z (blue), UP\_xy (red) and VL (orange) simulations. For each simulation outputs, two linear regressions have been conducted: one on the full East-Antarctic plateau dataset (thin lines) and one on the same dataset without the temperatures below $-43.5$$^{\circ}$C (bold lines) indicated by the vertical dashed line. The corresponding formulas of these latter are also shown.}
\label{fig3}
\end{figure}

We compare now our simulated spatial $\delta^{18}$O--temperature relationship and $\delta^{18}$O values for a given temperature to those from the data compiled by \citet{masson-delmotte2008}. Over the full temperature range, the spatial gradient is 0.83 \textperthousand.$^{\circ}$C$^{-1}$ in the VL simulation, very close of the observed one (0.80 \textperthousand.$^{\circ}$C$^{-1}$). We make the same comparison but by restricting the dataset to the ones on the East Antarctic plateau (Figure \ref{fig3}). As noticed previously, the average modeled temperature over Antarctica is overestimated whatever the simulation considered. Especially, no simulated temperature reaches a value below -50$^{\circ}$C. Yet, simulated $\delta^{18}$O values are too depleted for the temperature range between -49$^{\circ}$C and -43.5$^{\circ}$C.  As a consequence, a steeper modeled $\delta^{18}$O--temperature gradient is observed for the lowest temperatures, giving a modeled global gradient of 1.24 \textperthousand.$^{\circ}$C$^{-1}$ (thin orange line). If we restrict the fit to simulated temperatures higher than -43.5$^{\circ}$C (vertical dashed line in Figure \ref{fig3}), corresponding to the change of slope in the simulated $\delta^{18}$O--temperature relationship (thick orange line), the simulated gradient (0.96 \textperthousand.$^{\circ}$C$^{-1}$) is in more reasonable agreement with the one from the observations (0.85 \textperthousand.$^{\circ}$C$^{-1}$).

We now discuss possible reasons that could explain why the simulated $\delta^{18}$O--temperature slope is too steep at very low temperatures. First, it could be related to missing representation of fractionation during sublimation from the surface. As for all AGCMs equipped with water isotopes, fractionation at sublimation is not taken into account in LMDZ-iso. However, this effect would lead to further decrease of the water vapor $\delta^{18}$O in polar region and hence contribute to an even steeper $\delta^{18}$O--temperature slope at low temperature (hence further accentuate the mismatch). Second, the slope mismatch could be related to poorly represented kinetic fractionation. As in all the other models equipped with water isotopes, the parameterization of kinetic effect during vapor-to-solid condensation is represented empirically using a linear relationship between the supersaturation and the condensation temperature \citep{risi2010_lmdz}. A modification of the temperature can thus induce some change in the $\delta^{18}$O of the condensate, but this effect is of second order compared to the distillation effect explaining much of the slope between $\delta^{18}$O and surface temperature. Third, the slope mismatch could be related to a poor representation of the atmospheric boundary layer and of its related inversion temperature \citep{krinner1997, masson-delmotte2006}, as shown in the Figure \ref{fig2}. In LMDZ-iso, the warm bias in the simulated condensation temperature is smaller than that in the simulated surface temperature. Therefore, the water vapor masses continue to be distillated when moving away from the coast, while the cooling simulated at the surface from the coast to the remote region of the East Antarctic plateau is much less steep than in the reality.

\subsection{Comparison of the different simulations}
\subsubsection{Effects of the diffusive properties of the advection scheme}
We compare here the results from our different present-day simulations at a R96 grid resolution. The UP\_z simulation (upstream vertical advection, Figure \ref{fig1}b) increases the bias a little in $\delta^{18}$O, but its results stay relatively close of the $\delta^{18}$O values from the VL simulation, indicated by the similar average values that differ only by 0.89? for all Antarctica (Table \ref{table1}) that is smaller than the mean of the 60$^{\circ}$ S -- 90$^{\circ}$ S standard deviations. On the other hand, the $\delta^{18}$O outputs from the UP\_xy simulation (upstream horizontal advection, Figure \ref{fig1}c) display greater differences with the VL simulation ones, and so with the isotopic data, as revealed by the mean UP\_xy $-$ VL difference in $\delta^{18}$O of 4.31 \textperthousand. This is even more significant when focusing on the East Antarctic plateau, with a model--data difference in $\delta^{18}$O reaching 20 \textperthousand\ at some locations. The annual mean$\delta^{18}$O and $\delta$D values from the UP\_xy simulation are increased by 6.39 \textperthousand\ and 44.26 \textperthousand\ compared to the VL simulation average values, in less agreement with the observations as shown by their respective RMSE values (Table \ref{table1}, blue background). It shows that the diffusive property of the advection scheme on the horizontal plane is essential to better model the water isotope distribution, especially over Antarctica. To go further, one can also compare the $\delta^{18}$O values at a fixed temperature for the UP\_xy and VL simulations (Figure \ref{fig3}, red and orange crosses respectively). The $\delta^{18}$O in precipitation for a temperature of $-32$$^{\circ}$C over the East Antarctic plateau is already smaller by 6.1 \textperthousand\ in the VL simulation. This very significant difference in initial $\delta^{18}$O can be attributed to the proportion of mixing against distillation that affects the water vapor during its transport. This lends support to our hypothesis that too much diffusive mixing along the poleward transport leads to overestimated $\delta^{18}$O because dehydration by mixing follows a more enriched path than dehydration by Rayleigh distillation \citep{hendricks2000, galewsky2010}. We expect the relative contribution of mixing vs. distillation to have the largest impact on $\delta^{18}$O at latitudes where eddies are the most active. This is probably why the $\delta^{18}$O difference between the VL and UP\_xy simulation becomes large in mid-latitudes, over the austral ocean before arriving at the Antarctica coast (not shown), hence the difference in ``initial'' $\delta^{18}$O in Figure \ref{fig3}.

\begin{table*}[htbp]
\caption{Comparison of the observed annual mean values of temperature ($T$), $\delta^{18}$O and $\delta$D for the full Antarctic dataset with four different LMDZ-iso simulations, combining different horizontal resolutions (R96 and R144) and different advection schemes (UP\_xy and VL).}
\centering
\begin{tabular}{l | c | cccc}
\hline
							&	Mean	&	Mean UP\_xy	&	Mean VL	&	Mean UP\_xy	&	Mean VL	\\
							&	data		&	R96			&	R96		&	R144		&	R144	\\
\hline

$T$ ($^{\circ}$C)				&	$-36.93$	&	$-30.69$		&	$-31.54$	&	$-31.45$		&	$-32.10$	\\
$\delta^{18}$O (\textperthousand)	&	$-36.76$	&	$-31.43$		&	$-35.74$	&	$-33.85$		&	$-37.64$	\\
$\delta$D (\textperthousand)		&	$-289.62$	&	$-251.34$		&	$-279.49$	&	$-267.17$		&	$-289.55$	\\
\hline
\end{tabular}
\label{table2}
\end{table*}

As noticed in section \ref{section_temp}, all our simulations overestimate the average temperature in Antarctica and even more on the East Antarctic plateau. A more diffusive advection on the horizontal or on the vertical increases the mean temperature value by 0.85 and 1.03$^{\circ}$C respectively compared to the VL result. To explain such an influence of the advection on the temperature over Antarctica, even secondary, one can hypothesize that the Antarctic continent is better isolated, and so colder, when the advection of the model is less diffusive. If we focus now on the link between the temperature and the $\delta^{18}$O in precipitation over all the continent, the $\delta^{18}$O--temperature gradients according to our different R96 simulations UP\_z, UP\_xy and VL are at 0.79, 0.69 and 0.83 \textperthousand.$^{\circ}$C$^{-1}$ respectively. The difference between the VL and UP\_xy gradient shows an effect of diffusive properties of the large-scale transport on the distillation process. This difference between the modeled $\delta^{18}$O--temperature gradients is reduced if we restrict to the temperature range above $-43.5$$^{\circ}$C over the East Antarctic plateau, with gradients of 0.93 and 0.96 \textperthousand.$^{\circ}$C$^{-1}$ according to the UP\_xy and VL simulations respectively (Figure \ref{fig3}, red and orange thick lines). 

Since the simulated temperature difference between the UP\_xy and VL simulations is in the error margin, i.e.\ less than 1$^{\circ}$C, we do not expect that the temperature difference explains the difference in spatial slopes. Rather, the larger slope in simulation VL is due to the larger relative contribution of Rayleigh distillation compared to mixing.

\subsubsection{Effects of the horizontal grid resolution}
We test now the hypothesis that to increase the horizontal resolution is equivalent to using an advection scheme that is less diffusive. The Antarctica mean results are summarized in the Table \ref{table2}. Compared to the UP\_xy R96 simulation, the average value of $\delta^{18}$O in precipitation is decreased by 4.31 \textperthousand\ when the advection scheme is improved (VL R96), and by 2.42 \textperthousand\ when the horizontal resolution is increased (UP\_xy R144), in better agreement with the observations. The picture is the same for the $\delta$D outputs. The decrease of the mean modeled temperature values, smaller than the mean of the long-time standard deviations on the 60$^{\circ}$ S -- 90$^{\circ}$ S area, is the same by changing the advection scheme or by increasing the resolution: by 0.85$^{\circ}$C and 0.76$^{\circ}$C respectively. The best results are reached by improving both the advection scheme and the horizontal resolution at the same time, with model--data differences in temperature and $\delta^{18}$O of 4.83$^{\circ}$C and 0.88 \textperthousand\ respectively. This confirms that an increase of the horizontal resolution plays the same role as an improvement in the representation of the advection scheme on its horizontal plane \citep{hourdin1999}. It is worth mentioning that the improvement in model--data agreement using a higher horizontal grid resolution is probably not only due to an improved representation of the advection in the model but, among others, to a better resolved Antarctic topography and near surface circulation. Our results are consistent with the study of \citet{werner2011} that shows, for an increased horizontal resolution, a better agreement of the simulated isotopic delta values and of the water isotope--temperature gradient with the observations. The seasonality in modeled Antarctic temperature and precipitation, which influences the precipitation-weighted isotope values, is not altered by the change of advection scheme or horizontal grid resolution.

\subsubsection{Effects on the LGM to present-day change in temperature}
\begin{figure}[htb]
\center
\includegraphics[width=0.95\columnwidth]{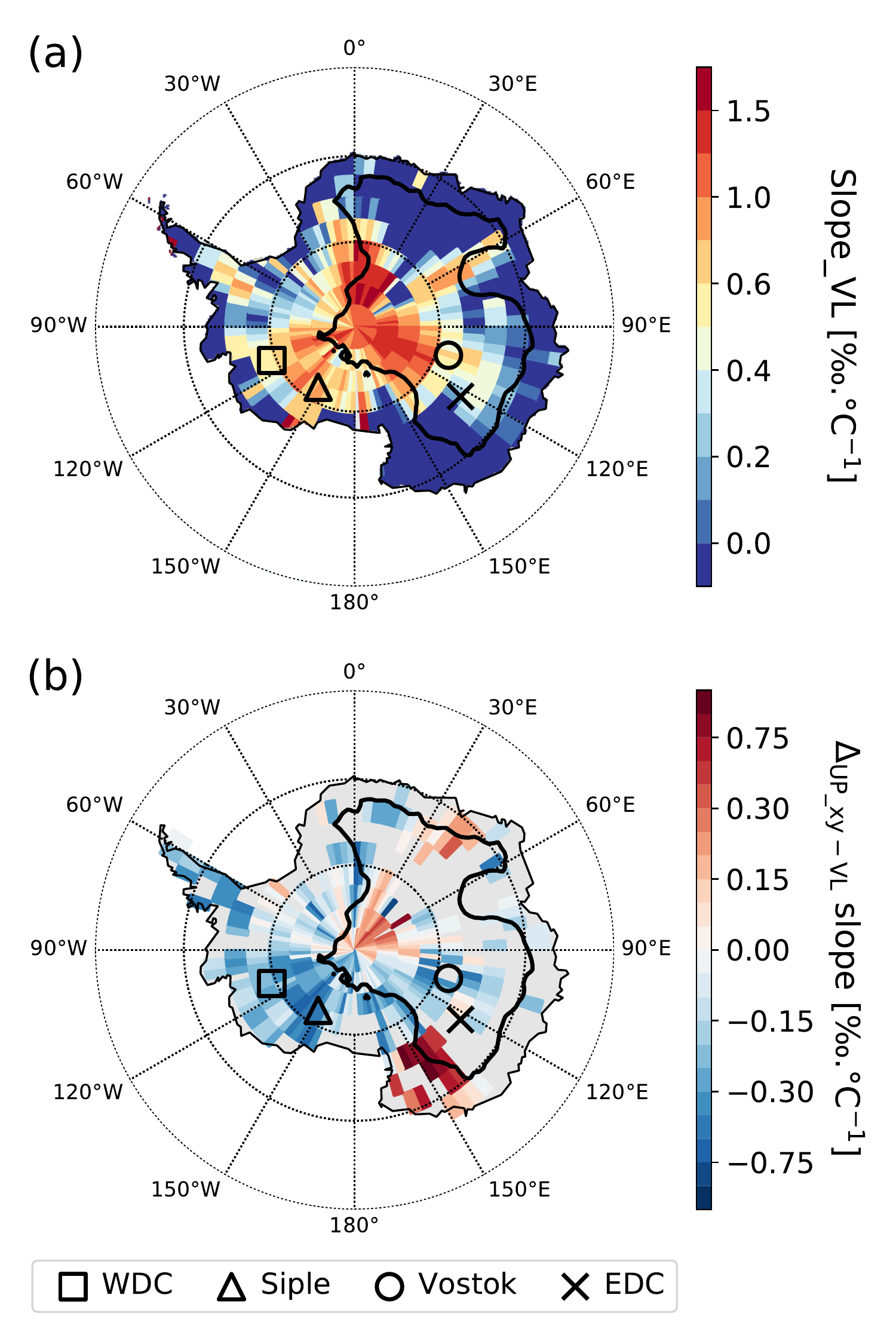} 
\caption{(a) Temporal slope ($\Delta\delta^{18}\textrm{O} / \Delta T$) between the present-day (PD) and LGM according to the VL simulations. (b) Difference with the temporal slope deduced from the UP\_xy simulations. The grey areas indicate where the LGM$-$PD difference in $\delta^{18}$O is lower than the standard deviation of the interannual variability. Square, triangle, circle and cross symbols represent the WDC, Siple Dome, Vostok and EDC sites respectively.}
\label{fig4}
\end{figure}

We evaluate now the effects of the diffusive properties of the advection scheme on the $\delta^{18}$O--temperature temporal slope at different Antarctic locations between LGM and present-day. For this, we compare the VL and UP\_xy LGM simulations with their present-day ``free'' counterparts (i.e.\ without nudging). Figure \ref{fig4}a shows VL simulated temporal slopes in $\delta^{18}$O--temperature at each location as $\Delta \delta^{18}$O$ / \Delta T$. Over East Antarctica, the simulated temporal slope becomes larger inland, going from near to 0 \textperthousand.$^{\circ}$C$^{-1}$ on the coast to 2 \textperthousand.$^{\circ}$C$^{-1}$ on the deep East Antarctic plateau. Over West Antarctica, the temporal slopes are more heterogeneous and show variations from 0.1 to 1 \textperthousand.$^{\circ}$C$^{-1}$. The slopes at WDC (Wais Divide Core, square symbol), Siple Dome (triangle symbol), Vostok (circle symbol) and EDC (cross symbol) are of 0.80, 0.66, 0.73 and 0.37 \textperthousand.$^{\circ}$C$^{-1}$ respectively (calculated by considering the 9 grid cells centered on each drill location). Figure \ref{fig4}b shows the differences between the UP\_xy and VL simulated slopes. As the spatial slopes from the VL simulation are higher than the UP\_xy one, it is expected that the LGM to present-day temporal slopes are larger in the VL simulation than in the UP\_xy one because of the colder temperatures and of the higher relative contribution of the Rayleigh distillation. This is the case on the western part of the continent where the UP\_xy temporal slopes are smaller than the VL ones by up to 0.6 \textperthousand.$^{\circ}$C$^{-1}$. On the East Antarctic plateau, the VL $\delta^{18}$O--temperature slopes are larger between the longitudes 100$^{\circ}$ E and 140$^{\circ}$ E than in UP\_xy as well. The slope values are smaller in VL than in UP\_xy only over the center of the continent, which corresponds to the lowest simulated temperature values and so where the gradient between the $\delta^{18}$O and temperature values is much steeper (see 2$^{\textrm{nd}}$ paragraph of section \ref{section_spatial_rel}), especially for LGM conditions.

We highlight now the consequences that an excessively diffusive horizontal advection may have on the LGM cooling reconstructed from simulated temporal slopes. For this, we calculated the difference in the LGM to present-day change in temperature deduced from the observed changes in $\delta^{18}$O and simulated slopes ($\Delta \delta^{18}$O$ / \Delta T$) by VL and UP\_xy, at WDC and Siple Dome stations for West Antarctica, and Vostok and EDC stations for East Antarctica (Table \ref{table3}). For WDC, Siple Dome and Vostok stations, the average changes in $\delta^{18}$O between LGM and present-day are larger in the VL simulation, $-11.98$, $-9.88$ and $-3.90$ \textperthousand\ respectively, than in the UP\_xy simulation ($-7.52$, $-6.24$ and $-2.07$ \textperthousand\ respectively). For the EDC station, the difference in LGM to present-day change in $\delta^{18}$O between our VL and UP\_xy simulations can be considered in the margin error ($-1.97$ and $-1.70$ \textperthousand\ for VL and UP\_xy simulations respectively). The values deduced from the VL simulation are also in better agreement with the observations (Table \ref{table3}). As previously noticed, the temporal $\delta^{18}$O--temperature gradients at WDC, Siple Dome, Vostok and EDC are also larger in the VL simulation by 57 \%, 65 \%, 92 \% and 26 \% respectively, compared to the values deduced from the UP\_xy simulation. As a consequence, if one applies the slope from the UP\_xy simulation instead of the VL one, we would overestimate the present-day-to-LGM cooling at these stations by 4.34, 5.83, 3.76 and 2.95$^{\circ}$C respectively. Even if the change in the temporal $\delta^{18}$O--temperature gradient at EDC is relatively smaller than for the other stations, the consequence on the reconstructed present-day-to-LGM cooling value is still important. These results show that a more diffusive advection scheme on the horizontal plane can affect greatly the LGM to present-day temperature change deduced from observed $\delta^{18}$O in precipitation.

\begin{table*}[htbp]
\caption{Sites of interest and their geographical coordinates, observed LGM to present-day changes in $\delta^{18}$O (Wais Divide: \citet{wais2013}, \citet{schoenemann2014}; Siple Dome: \citet{brook2005}, \citet{schoenemann2014}; Vostok: \citet{vimeux2001}, \citet{landais2008, landais2012}; EDC: \citet{epica2004}, \citet{stenni2010}), simulated LGM to present-day changes in $\delta^{18}$O (UP\_xy and VL simulations), simulated temporal $\delta^{18}$O--temperature slopes and VL$-$UP\_xy difference in reconstructed temperature change deduced from the observed $\delta^{18}$O changes and the simulated temporal slopes. This difference is calculated as $- (slope\_UP\_xy - slope\_VL) \times \Delta\delta^{18}O_{\textrm{obs}} / slope\_VL^2$. Negative signs indicate that when using the UP\_xy temporal slope, we overestimate the LGM cooling compared to using the VL temporal slope. The model values are based on the spatial averages over the 9 grid cells surrounding the ice cores geographical coordinates.}
\centering
\begin{footnotesize}
\begin{tabular}{lcccccccc}
\hline
Site			&	Latitude	&	Longitude	&	LGM$-$PD					&	LGM$-$PD					& LGM$-$PD					&	Slope							&	Slope							&	Difference in			\\
			&			&			&	$\delta^{18}$O$_{\textrm{obs}}$	&	$\delta^{18}$O$_{\textrm{UP\_xy}}$	& $\delta^{18}$O$_{\textrm{VL}}$	&	UP\_xy							&	VL								&	reconstructed			\\
			&			&			&	(\textperthousand)				&	(\textperthousand)				& (\textperthousand)				&	(\textperthousand.$^{\circ}$C$^{-1}$)		&	(\textperthousand.$^{\circ}$C$^{-1}$)		&	$\Delta T$ ($^{\circ}$C)	\\
\hline
WDC		&	$-79.47$	&	$-112.08$	&	$-9.44$	&	$-7.52$	&	$-11.98$	&	$0.51$	&	$0.80$	&	$-4.34$	\\
Siple Dome	&	$-81.67$	&	$-148.82$	&	$-9.8$	&	$-6.24$	&	$-9.88$	&	$0.40$	&	$0.66$	&	$-5.83$	\\
Vostok		&	$-78.47$	&	$106.87$	&	$-5.66$	&	$-2.07$	&	$-3.90$	&	$0.38$	&	$0.73$	&	$-3.76$	\\
EDC			&	$-75.10$	&	$123.35$	&	$-6.74$	&	$-1.70$	&	$-1.97$	&	$0.31$	&	$0.37$	&	$-2.95$	\\
\hline
\end{tabular}
\end{footnotesize}
\label{table3}
\end{table*}

\section{Conclusions}
We have tested with LMDZ-iso if the warm and isotopically enriched biases in Antarctica, frequently observed in the AGCMs, are associated with the diffusive property of the advection scheme. The simulated water isotope contents in Antarctica are very sensitive to the diffusive character of the water vapor transport on the horizontal plane. The higher the contribution of mixing (i.e.\ diffusion), the more enriched the precipitation. These findings are even more striking for the East Antarctic plateau where the main ice cores allowing paleoclimate reconstructions are located. Moreover, because the diffusive character of the large-scale transport influences the temperature in this region, even in a light way, this has an impact on the modeled water isotopic composition through the Rayleigh distillation. So, we conclude here that the excessive numerical diffusion has a large influence on the enriched isotopic bias. For the spatial isotope--temperature relationship over the East Antarctic plateau observed in LMDZ-iso, this latter is improved for the temperatures above $-43.5$$^{\circ}$C, in more reasonable agreement with the observations. At the lowest temperatures (i.e.\ still over the East Antarctic plateau), that the model is not able to reach, the non-linearity observed in our simulations (the spatial $\delta^{18}$O--temperature relationship is steeper for the lowest temperatures) can be unlikely explained at first order to missing or poorly represented kinetic fractionation. One can speculate that the water masses continue to be distillated when moving away from the coast, hence depleting the water vapor in heavy isotopes while the modeled temperature decrease from the coast to the remote region of the East Antarctic plateau is much less steep than in the reality. This more pronounced effect in the case of a more diffusive horizontal advection scheme could be due to the deteriorated representation of the inversion temperature. The temporal isotope--temperature relationship at some locations in Antarctica can be influenced by the diffusive properties of the advection scheme on its horizontal domain. As for the spatial gradient, an excessive numerical diffusion has the consequence to decrease the isotope-temperature temporal gradient, leading to a wrong estimation of the LGM to present-day temperature change deduced from observed $\delta^{18}$O. Our study demonstrates that a representation of the advection scheme in the AGCMs taking into account water isotopes and isotopic gradients, especially on the horizontal domain, is an important step toward a more realistic modeling of water isotopes over Antarctica. Another way to improve this aspect is to increase the spatial resolution, which has the same effect as applying a less diffusive advection scheme on the water isotopic composition and the temperature. This study shows again the importance of using water stable isotopes in GCMs for the evaluation and quantification of the processes influencing the hydrological cycle, including the advection of water vapor. We expect our main results (excessive diffusive advection leading to warmer temperatures, moister boundary layer and more enriched water vapor) to be robust and to hold in other models as well. However, the quantitative response to the advection scheme may be modulated by the representation of boundary layer processes and high-cloud microphysics in each model.

\section*{Acknowledgements}
We thank A. Landais for her useful suggestions on this manuscript, J.-B. Madeleine and C. Genthon for their kind support about inversion temperatures over EDC, and R. Guzman, C. Listowski and H. Chepfer for their help with the GOCCP data. This work was granted access to the HPC resources of IDRIS under the allocation 0292 made by GENCI. The research leading to these results has received funding from the European Research Council under the European Union's Seventh Framework Programme (FP7/20072013)/ERC grant agreement no. 30604. We thank Christophe Genthon and the CALVA program for acquiring and distributing meteorological data at Dome C (\url{http://www.lmd.jussieu.fr/~cgenthon/SiteCALVA/CalvaBackground.html}) with the support of the french polar institute (IPEV). We also acknowledge the WCRP-BSRN network and Angelo Lupi for the dissemination of radiation data. Radiosoundings data at Dome C are freely distributed in the framework of the IPEV/PNRA Project ``Routine Meteorological Observation at Station Concordia'' --- \url{www.climantartide.it}.

\end{document}